# Advanced drag-free concepts for future space-based interferometers: acceleration noise performance

**D Gerardi[1], G Allen[2], J W Conklin[2], K-X Sun[2], D DeBra[2], S Buchman[2], P Gath[1], W Fichter[3], R L Byer[2], and U Johann[1]**

[1] Astrium Satellites, Friedrichshafen, Germany

[2] Hansen Experimental Physics Laboratory, Stanford University, Stanford CA

[3] Institute of Flight Mechanics and Control, University of Stuttgart, Stuttgart, Germany

E-mail: domenico.gerardi@astrium.eads.net

**Abstract.** Future drag-free missions for space-based experiments in gravitational physics require a Gravitational Reference Sensor with extremely demanding sensing and disturbance reduction requirements. A configuration with two cubical sensors is the current baseline for the Laser Interferometer Space Antenna (LISA) and has reached a high level of maturity. Nevertheless, several promising concepts have been proposed with potential applications beyond LISA and are currently investigated at HEPL, Stanford, and EADS Astrium, Germany. The general motivation is to exploit the possibility of achieving improved disturbance reduction, and ultimately understand how low acceleration noise can be pushed with a realistic design for future mission.

In this paper, we discuss disturbance reduction requirements for LISA and beyond, describe four different payload concepts, compare expected strain sensitivities in the "low-frequency" region of the frequency spectrum, dominated by acceleration noise, and ultimately discuss advantages and disadvantages of each of those concepts in achieving disturbance reduction for space-based detectors beyond LISA.

## 1. Introduction

The technology for following a geodesic in space was first suggested by Lange [1] and DeBra [2], and demonstrated by the flight of the TRIAD [3] satellite using the DISCOS system. A free-floating test mass (TM), located near the center of mass of the spacecraft (SC), is kept centered in its housing by adjusting the SC position using thrusters. A control loop provides as outputs the commands for the SC thrusters, using the TM position in its housing as inputs. GP-B [4] was the first experiment to demonstrate the drag-free control in the accelerometer mode, viz. the gyro suspension system effort was used as the error signal for drag free control of the spacecraft.

Inertial navigation systems are historically divided in two categories: drag-free sensors with TM position controlled by SC thrusters, and accelerometers with TM position controlled by forcing provided from its own housing.

*The drag-free mode:* As described above the SC in the drag-free mode follows the free-floating TM with no forces applied to the TM. Drag-free is typically implemented in one of three control approaches:

1) Pure drag-free in which the control loop has as inputs the position of the TM in its housing and as output the commands for the SC thrusters. The position and attitude of the TM is kept stable in the housing by flying the SC around the TM. Examples of such missions are TRIAD, and GP-B.

2) Accelerometer-mode I drag-free in which the control loop has as inputs the control effort signals of the TM inner control loop in its housing and as output the commands for the SC thrusters. The control law minimizes the control effort on TM at a nominal location.

3) Accelerometer-mode II drag-free in which the slower outer control loop has as inputs the control effort signals of the TM in its housing and as output the commands for the SC thrusters, as in accelerometer-mode I drag-free. The inner, faster control loop of the TM in its housing uses the TM position as inputs to perform the fine adjustments of the TM position using the housing-based forcing system: DRS/LPF [6], LISA [7,8] STEP [9], GP-B. This mode can perform at close to the level of the pure drag-free mode.

*The accelerometer mode:* In this mode the position of the TM in its housing is kept fixed by a forcing control loop with the control effort as the input, with the output force applied to the TM housing forcing system (and no forcing of the SC). GRACE [10] and CACTUS [5] are examples of such missions.

Note that the drag-free mode is optimized by minimizing the TM-to-housing stiffness, while the accelerometer mode requires maximizing it.

In this paper, we mainly refer to LISA as being the most demanding drag-free system envisioned by an active space program, but discuss aspects of disturbance reduction systems that can largely and generally apply to future drag-free missions beyond LISA. We discuss disturbance reduction requirements for LISA and beyond, and perform a trade-study of four different configurations. With reference to possible applications to missions beyond LISA, the trade study in this paper will concentrate on the limits of the acceleration noise on the TM and only refer to general considerations and conclusions regarding: complexity of implementation, simplicity of initialization in space and operations, overall reliability.

## 2. Disturbance reduction for LISA and beyond

LISA will be a space-based observatory of gravitational waves, which will operate in a frequency range from $10^{-4}$ Hz to 1 Hz. To detect and observe gravitational radiation, LISA will require:

- free-falling objects, i.e. test masses evolving in a nearly perfect geodetic motion, shielded from external disturbances by a drag-free SC;
- measurements of any fluctuation in the relative distance among free-falling objects through a very long baseline laser interferometer in space.

The concept features a constellation of three spacecrafts shielding the test masses from external disturbances in a drag-free mode, and flying in a one-year solar orbit, in a nearly equilateral triangle at 60° to the ecliptic plane; the center of the triangle is 20° behind the Earth; arm-lengths are nominally 1/30 of an AU ($l \approx 5 \cdot 10^9$ m)[4]. The constellation rotates in its plane once per year, and the constellation plane itself precesses around the ecliptic pole once per year.

The strain sensitivity which is targeted for LISA is addressed in several publications; the constellation aims at detecting gravitational radiation, and characterize GW sources with a strain sensitivity of $1.1 \cdot 10^{-20}$ Hz$^{-1/2}$ (at 5 mHz), ideally [19] rising as $f^{-2}$ down to lower frequencies of $10^{-4}$ Hz[5].

---

[4]slightly changing over one-year period, due to orbital mechanics design, which would cause the constellation angle to change by approximately ±1° over the year.

[5] $f^2$ is a purely theoretical behaviour for the sensitivity curve down to $10^{-4}$ Hz; in practice, low-frequency thermal fluctuations, fluctuations in the interplanetary magnetic field coupled to DC gradients of local (on-board) magnetic

Disturbances acting on the test masses that would cause acceleration noise and perturb their motion, are the fundamental limit to instrument sensitivity at low frequency, below approximately 5 mHz . Errors in the measurement of any fluctuation in the relative distance between the test masses would drive strain sensitivity at higher frequencies, above 5 mHz.

Below 5 mHz, the strain sensitivity $\delta h/h$, on one single link of the LISA constellation, can be written as:

$$\frac{\delta h}{h} = T_{GW}^{TDI} \cdot \frac{\delta l}{l} \approx \frac{k_{av}}{\sin\alpha \cdot \mathrm{sinc}\left(\frac{2l}{c}f\right)} \cdot \frac{\delta l}{l} \qquad (1)$$

i.e. it is given by the TDI transfer function[6], $T_{GW}^{TDI} = k_{av}\left[\sin\alpha \cdot \mathrm{sinc}\left(\frac{2l}{c}f\right)\right]^{-1}$, times the relative fluctuation of the arm-length $\delta l/l$; $k_{av}$ is a factor that would result from averaging GW signals over different directions in the sky, and different polarizations; $\alpha$ is the constellation angle, $l$ the arm-length, $f$ the frequency of evaluation, $c$ the speed of light.

Since disturbances $\delta a$ acting on each test mass, and measurement errors $\delta l_{meas}$ will make the relative distance between two masses on a single link fluctuate as:

$$\frac{\delta l}{l} = \frac{1}{l}\sqrt{(\delta l_{meas})^2 + \left(\frac{\delta a}{(2\pi f)^2}\right)^2} \qquad (2)$$

the sensitivity of the instrument, $\delta h/h$, on a single link at low frequency can be expressed as:

$$\frac{\delta h}{h} \approx \frac{k_{av}}{\sin\alpha \cdot \mathrm{sinc}\left(\frac{2l}{c}f\right)} \cdot \frac{1}{l}\sqrt{(\delta l_{meas})^2 + \left(\frac{\delta a}{(2\pi f)^2}\right)^2} \qquad (3)$$

As a result, if a sensitivity of $1.1 \cdot 10^{-20}$ Hz$^{-1/2}$ at 5 mHz has to be achieved, with $l = 5 \cdot 10^9$ m arm-length, in the presence of a measurement noise that can be as large as 40 pm/√Hz, then acceleration disturbances acting on each test mass should be kept below $\delta a < 3 \cdot 10^{-15}$ m s$^{-2}$/√Hz.

---

field, residual accumulating charges on the TM would be the sources of disturbances that would cause the sensitivity curve to rise more steeply than $f^2$ down to $10^{-4}$ Hz; making the design less vulnerable to these three sources of disturbances would allow for approaching the ideal $f^2$ behaviour; reducing the deviation from the ideal sensitivity at low frequency by careful design would allow for: observing MBH binaries with ~$10^7$ solar masses, a better distance and angular resolution of GW sources, mapping dark energy very far back in time when combining GW data with EM observations.

[6] conventional methods for suppressing laser frequency noise in equal-arm interferometers cannot apply to LISA, due to un-avoidable differences in arm-lengths that cause laser phase noise to experience different delays in different arms; a technique (Time Delay Interferometry, TDI) relying on time-shifting and linearly combining independent measurements to suppress laser phase noise for LISA during on-ground data analysis has been proposed at JPL; an overview of its theory, and mathematical foundations can be found in [44].

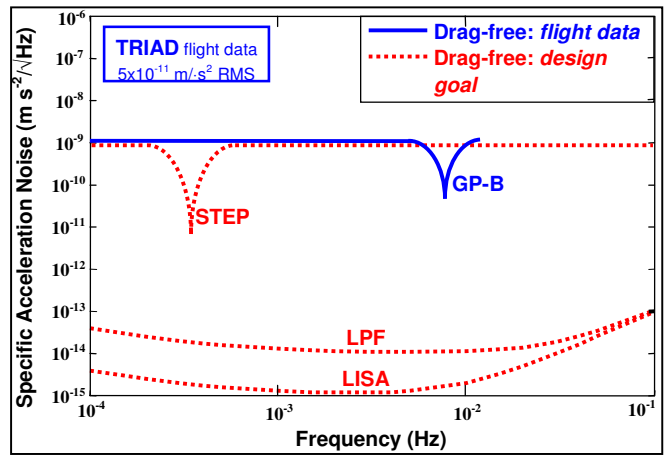

**Figure 1.** Disturbance reduction requirements for some modern drag-free missions.

Figure 1 shows the frequency spectrum requirements of the specific acceleration noise for a number of modern drag-free experiments: GP-B, STEP, DRS / LPF, and LISA. The best performances demonstrated to date are: the DISCOS instrument on the TRIAD flight with a specific acceleration noise of $5 \cdot 10^{-11}$ m·s$^{-2}$ RMS for an integration time of 3 days (4 µHz), and the GP-B drag-free sensor (one of four gyroscopes) with a specific acceleration noise floor of $1.2 \cdot 10^{-9}$ m s$^{-2}$ Hz$^{-1/2}$ in inertial space from < 0.01 mHz to 10 mHz [20] and a residual acceleration noise of $10^{-11}$ m s$^{-2}$ Hz$^{-1/2}$ at 12.9 mHz (1/77.5 s roll rate).

For the wide-band Laser Interferometer Space Antenna mission, the minimum improvement required over demonstrated performance is hence five to six orders of magnitude (figure 1). Even more demanding disturbance reduction would be required for missions beyond LISA.

The following questions, therefore, arise: how do we design to meet these extremely demanding requirements, and, ultimately, how low can local disturbances, other than large-scale gravitation, be pushed for a test mass in a drag-free SC with a realistic design?

In the attempt to address these questions, some of the main design choices that constitute the top-level design matrix for LISA are presented; four different payload configurations corresponding to different top-level design decisions are generated; acceleration disturbances expected for those configurations are estimated and compared; main choices that would drive strain sensitivity down at low frequency are identified and emphasized; guidelines for achieving disturbance reduction for missions beyond LISA are derived.

## 3. General design considerations and options

Figure 2 summarizes some design choices that form the top-level design matrix for the LISA payload. The hexagonal outlines represent the spacecraft with the two telescopes for the LISA configuration. The proof-mass lies within a reference housing. The received light is measured with respect to the housing, while a separate interferometer measures the displacement between the housing and the proof-mass.

Depending on which design decisions are taken at this level, different configurations result. The following four options are considered:

1) Two cubical test masses per SC and movable telescope assembly. This is the current baseline design for LISA: two cubical sensors per SC, a combination of capacitive sensing and optical sensing (optical sensing only in the direction of the two lines of sight of the LISA constellation) to measure relative position and attitude of the test masses with respect to the SC for drag-free control, a moving telescope assembly to cope with the annual variation of the in-plane constellation angle. No electrostatic actuation would occur in the direction of the two lines of sight during science runs, only a "weak" electrostatic suspension on non-drag-free degrees of freedom would be performed to keep two cubical test masses centered and aligned with respect to the their housings and with respect to the interferometers; SC attitude would be controlled based on inter-SC Differential Wavefront Sensing (DWS) signals.

| | Configuration 1 (LISA baseline) | Configuration 2 | Configuration 3 (MGRS) | Configuration 4 | |
|---|---|---|---|---|---|
| | 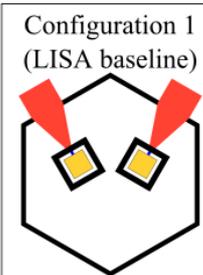 | 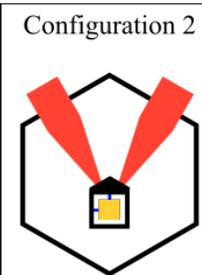 | 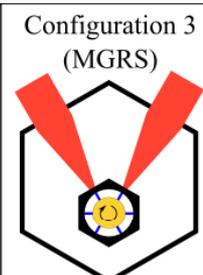 | 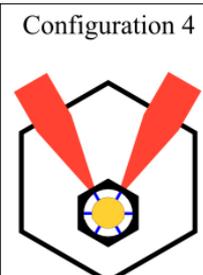 | Number of GRS |
| Number of PM per S/C | Two | One | One | One | |
| PM Shape | Cubes | Cube | Sphere (spinning) | Sphere (slowly rolling) | GRS design concept |
| PM Sensing System | Capacitive + Optical | Optical | Optical | Optical | |
| PM Suspension System | Electrostatic | Electrostatic (orientation only) | None | None | |
| Telescope pointing | Moving telescope assembly | Moving optical elements | Moving optical elements | Moving optical elements | |

**Figure 2.** Top-level design choices and resulting configurations of the LISA payload. The hexagonal outlines represent the spacecraft with the two telescopes for the LISA configuration, and the proof-mass is contained in its reference housing.

2) One cubical test mass per SC and In-Field-of-View pointing. This design features only one active Gravitational Reference Sensor (GRS) per SC, full optical sensing during science runs, movable mirrors inside the telescope assembly (In-Field-of-View pointing) that would allow for coping with the annual change in the constellation angle. With only one (cubical) TM per SC, no translational forcing but only attitude suspension of the TM would be required during science runs.

3) One (spinning) spherical test mass per SC with full optical sensing and In-Field-of-View pointing. The test mass would be spinning orthogonally to the ecliptic plane at a proper frequency (10 Hz) above the LISA band to spectrally shift the out-of-roundness and mass center offset errors that would result from imperfections in the manufacturing process of the sphere[7]; each of the two telescope assemblies would be fixed to the SC structure and movable mirrors inside the telescope assembly would be used as in configuration 2. No electrostatic suspension (forcing) would be applied to the test mass during science runs, the relative position of the sphere would be

---

[7] out-of-roundness errors might be as large as 20 $nm_{rms}$ according to state-of-the-art manufacturing techniques [21]; in-homogeneities of density for materials proposed for LISA test masses could cause a mass center offset ~ 1 μm if not properly controlled; combining measurements and proper fabrication processes allows for reducing the mass center offset to within 300 nm with potential of achieving 100 nm [45].

>sensed optically, and the position of the SC would follow (drag-free controlled) through µN−thrusters; SC attitude would be controlled based on inter-SC DWS signals.
>
>4) One (non-spinning) spherical test mass per SC with full optical sensing and In-Field-of-View pointing as in configurations 2, and 3. The sphere would not be spinning during science runs; after release it would be freely tumbling inside the sensor; optical sensing degraded by un-roundness and mass center errors would be used to make the SC follow the sphere in a drag-free mode; the position of the center of mass of the sphere would be reconstructed on-ground during data analysis based on high precision mapping of TM spherical harmonics [23].

The drag-free control for a spherical TM (first proposed with application to the LISA mission by [11]) in a LISA-like noise environment has been designed in detail in [23]; optimized controllers can be selected to allow for both "fast-spin" (TM spinning at a proper frequency above the upper bound of the LISA band, i.e. configuration 3), and "no-spin" operations (configuration 4).

*3.1. Gravitational Reference Sensors*

*3.1.1. LISA baseline GRS (Configuration 1).* The sensor envisioned for the current payload baseline is being developed and tested as a part of the LISA Technology Package (LTP) by a consortium of seven European space agencies and ESA with EADS Astrium Germany acting as Industrial Architect. The LISA Technology Package, [24], [25], will fly on board the LISA Pathfinder mission [26], [27]. LISA Pathfinder is currently in its implementation phase, and the core instrument (LTP) has passed the Critical Design Review.

The gravitational reference sensor is a capacitive sensor that features a cubical TM made of an alloy of Au-Pt, and surrounded by a configuration of gold-coated electrodes for capacitive sensing and electrostatic actuation. Sensor design, as well as experimental set-up for on-ground performance tests, is described in detail in several publications [24], [27].

For drag-free control proof-mass position and orientation are read-out out by a combination of capacitive and optical sensors. Optical sensors are used along the interferometer axis and to measure two-axis tilt angles of the TM ($x$, $\eta$, $\phi$). All other degrees-of-freedom are sensed by capacitive sensing with a 2 mm gap (increased to 4 mm along the interferometer axis).

For configuration 1, each arm contains an independent GRS, and electrostatic control forces are applied to control the position and orientation of each proof-mass. No control forces are applied along the interferometer lines of sight and along the $z$-axis of one proof-mass.

*3.1.2. Configuration 2.* Configuration 2 uses a modified variant of the LISA baseline GRS, fully instrumented with optical sensing for all degrees-of-freedom. The effective motion along the 60° interferometer lines of sight is synthesized based on the measured displacements of the optical sensors. The electrostatic suspension is used to control the orientation of the proof-mass, while the translational control effort is handled by the spacecraft.

3.1.3. *Modular Gravitational Reference Sensor (MGRS).*  To achieve drag-free operation and eliminate TM forcing, a non-supported spinning spherical TM with a housing that separates the main interferometer beam from the local TM measuring beams (the Modular Gravitational Reference Sensor, MGRS [14], [15], [16]) has been proposed and designed at Stanford, particularly with applications to future space-based detectors beyond LISA. This design achieves the following desirable characteristics for the drag-free sensor:

   a) a spherical TM does not require position and orientation forcing and uses only optical sensing, therefore minimizing disturbance forces and noise;
   b) no position forcing, and optical position sensing allows maximizing the TM-to-housing gap for reduction of the patch-effect by $d^{-3}$;
   c) TM spinning spectrally shifts the position measurement noise to above the science band;
   d) separating main interferometer and local TM measurement beams allows for high-intensity main beams and modular engineering.

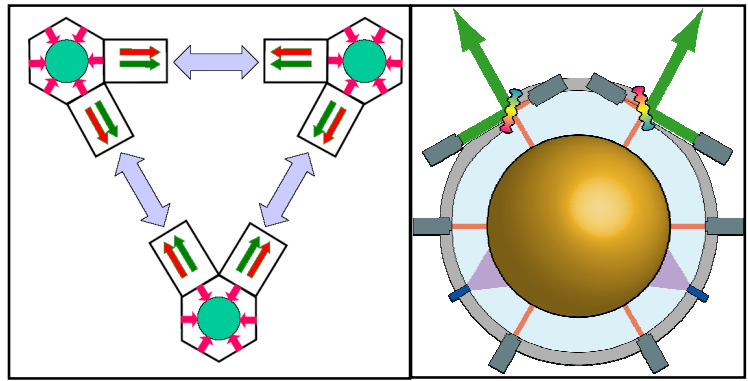

**Figure 3.** a), left, LISA schematic with 3 MGRS; b) schematic of MGRS design.

Figure 3a shows schematically the three MGRS (single-sphere with optical readout), as well as the three interferometer arms. Note the two colours for the beams, representing a double interferometer with a near-infrared Nd:YAG line and a green doubled-frequency one. The precision of the measurement scales as $\lambda^{-3/2}$, a factor of three better for the green line. Figure 3b shows a schematic of the MGRS design.
The proposed gap between test mass and housing is $d = 20$ mm. This gap would reduce the GP-B time-independent patch effects (GP-B gap size $d = 20$ µm), scaling as $d^3$, by about nine orders of magnitude, bringing it in line with LISA requirements.

Two factors need to be considered regarding time-independent forces: a) true time-independent forces are tolerable at much higher levels than those shown in figure 1, and b) other disturbance forces are likely to occur over these many orders of magnitude.  The time-dependence of the patch-effects is only beginning to be understood, and is clearly non-zero [11]. The time-dependent charging of the TM will give rise to cross-terms with the patches providing a time dependent patch mechanism even in the presence of absolutely constant patch-effects.  Furthermore, time-dependent patch effects only scale as $d^{-2}$, giving rise to a larger disturbance factor.  Other various disturbance mechanisms need to be reduced on their own if the overall performance is to meet LISA requirements.  To conclude, a gap as large as practical is highly desirable to meet LISA-level drag-free requirements.

Spinning the TM, at about 10 Hz, spectrally shifts asphericity errors above the upper bound of the science band, 1 Hz for LISA. The TM is designed with one larger moment of inertia, $I_1 = I_2 \approx 0.9 I_3$, which can be obtained by hollowing out portions of the sphere during the manufacturing process. This partially hollowed-out test mass design with $\Delta I/I = (I_3 - I_2)/I_1 \geq 0.1$ meets the following three goals: a) increase the frequency of the polhode motion $f_p \sim (\Delta I/I) f_s \approx 1$ Hz above the LISA bandwidth; b) allow fast damping of the polhode motion[8]; c) enable the mass center to be moved toward the geometric center through an iterative process whereby the mass center location is measured, material is removed from the heavy side of the sphere, then the sphere is re-rounded by lapping and polishing. A double-sided grating on the sensor housing isolates the GRS from external measurements. The distance from the grating to the TM center of mass is then measured, making the GRS modular, and thus insulating it from other experimental systems [29].

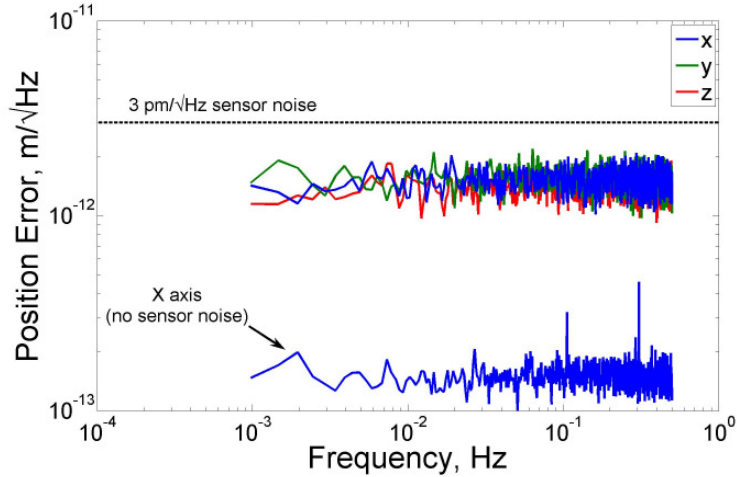

**Figure 4.** Simulation of residual displacement error of the MGRS using optical sensing.

The position of the TM center of mass is determined by mapping its surface with a number of optical beams (see below). With the center of mass measured optically, no active electrostatic systems are present to disturb the TM. Optical sensing offers a high-resolution method of sensing across a large gap while maintaining low disturbances. The sensing element is a low-finesse Fabry-Perot cavity formed between a Littrow mounted 900 lines/mm diffraction-grating and the surface of the proof-mass [30]. A three dimensional numerical simulation of a sphere spinning at 10 Hz has been performed to demonstrate the feasibility of measuring the TM center of mass [31], [32]. Six to eighteen optical sensors, with two sensors in the 60 degrees configuration of the LISA gratings are being used in the model. The position noise and surface roughness of the sphere are 50 nm. Figure 4 shows simulation results indicating that the TM center of mass can be measured to 3 pm/√Hz.

---

[8] with the fundamental polhode frequency above 1 Hz, it would still be possible to generate small error signals in the LISA band through the interaction of the slowly decaying polhode angle with high-order harmonics of the sphere's shape which are symmetric about the $I_3$ axis which do not get averaged out by the TM spin. To ensure that the effect of the decaying polhode does not corrupt the science signal it would be necessary to either actively damp out the polhode or to wait for it to damp out naturally as a result of internal energy loss [31]. Active damping can be accomplished by applying external torques using eddy currents [49].

3.1.4. *Configuration 4.* Configuration 4 uses a design similar to the MGRS, except the proof-mass does not have any moment of inertia difference ratio ($\Delta I/I$) tuning and it is not intentionally spun up. The sphere will then likely be slowly rotating as a result of residual torques when it is uncaged. A series of optical sensors records the proof-mass motion and the surface profile of the sphere. When the optical sensor data is sent to the ground it is compared with high resolution maps of the sphere's surface and a mechanics model. The high resolution maps are then used to subtract the sphere's surface roughness from optical signals to obtain an accurate measurement of the proof-mass' position.

Electrostatic patch effect induced forces may put a lower bound on spinning rate for the fast spin option, and an upper bound for the slow spin option. A charged patch on the sphere will interact with any charged patches on the housing producing a force at the spin frequency and its harmonics. The large gap will significantly reduce this effect for patches that are approximately 10 times smaller than the gap size ($l_{patch} \sim d/10$). For a sphere of radius $r = 2.85$ cm, this corresponds to about the 100$^{th}$ spherical harmonic. For the non-spinning case, we require that the sphere has a residual rotation rate of less than $10^{-6}$ Hz, so that disturbances due to patch effects occur at frequencies below $10^{-4}$ Hz. Since this effect occurs only at harmonics of the spin frequency, it is not included in this analysis, but does represent a significant potential noise source if the spin frequency were to exceed $10^{-6}$ Hz.

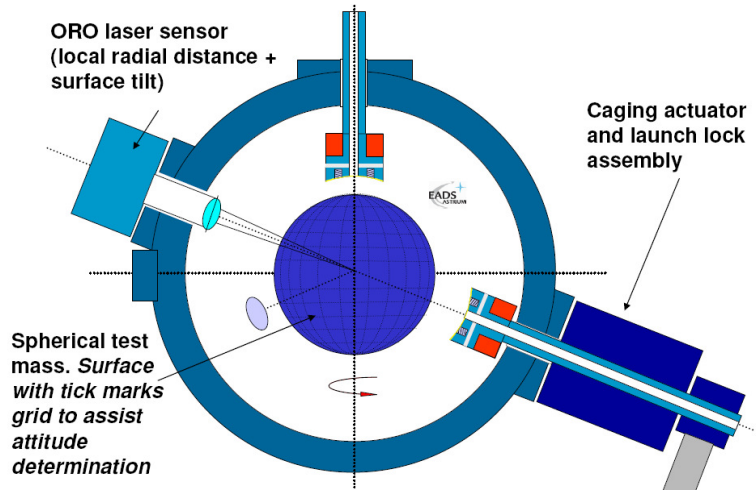

**Figure 5.** Schematic of GRS design in configuration 4, [12], [18], [48].

The in-flight position accuracy is limited by the sphere's out-of-roundness, resulting in jitter motion of 12 nm/√Hz, compared with sub-nanometer jitter for all other configurations. The increased gap size lowers the proof-mass to spacecraft stiffness, allowing the jitter requirement to be significantly relaxed compared with configurations 1 and 2.

3.2. *Relevant performance parameters*

Table 1 summarizes some of the main design parameters which are relevant to estimate acceleration noise for the four configurations; highlighted sections are where differences exist between the cubic and spherical GRS designs. For the GRS in configurations 1 and 2, the proof masses are nominally 1.9 kg cubical alloys of Au-Pt (73% - 27%), whose diamagnetic and paramagnetic properties combine to give low magnetic susceptibility $\chi = 1.7 \cdot 10^{-5}$, and high density $\rho_{TM} = 1.96 \cdot 10^4$ kg·m$^{-3}$.

For the spherical GRS in configurations 3 and 4, the same inertial properties (mass), and same material[9] were kept as for the baseline TM, setting the radius of the sphere to $r = 2.85$ cm, which puts no particularly demanding requirements on the design of the optics and the GRS housing. The choice of full optical sensing and no electrostatic forcing allows a large gap ($d = 20$ mm in the remainder of the paper) between the sphere and the housing walls, which reduces stiffness coupling $k$ of the TM to the SC (section 4), and makes the sensor less vulnerable to charge fluctuations (coupled to DC voltages in the housing) and temporal variations of effective potential due to patch fields (section 4). For the cubical GRS in configurations 1 and 2, a gap of $d = 4$ mm between the TM and its housing is assumed.

TM magnetic properties, such as susceptibility to magnetic fields, are assumed to be identical in all four configurations; fluctuations of local magnetic field, and local magnetic field gradient as summarized in table 1, are considered to be realistically achievable in all configurations with a suitable magnetic design of the surrounding SC, and a proper design / allocation of active components on-board.

An average interplanetary magnetic field of $3 \cdot 10^{-8}$ T is weaker than the local field, but has fluctuations of single components as large as $\delta B_{int} = 40$ nT/√Hz at 1 mHz, rising as steeply as $1/f$ at lower frequencies (from Ulysses data according to [13]). Furthermore, if a Poisson-distributed charging rate of $\dot{q} = 260$ e/s is assumed, fluctuations in the charge on the proof mass as large as $\delta q \sim (2e\dot{q})^{1/2} / 2\pi f \sim 5.85 \cdot 10^{-16} (1 \text{ mHz}/f)$ C/√Hz would be expected for all configurations in the absence of any active charge control.

Relevant design parameters for the capacitive sensor adopted for the baseline GRS are summarized in table 4; fluctuations in voltage difference inside the sensor of 10 µV/√Hz are assumed.

A residual gas pressure of $p = 10$ µPa inside TM housing, and a DC temperature of $T = 293$ K, with fluctuations of temperature differences across the housing as large as $\delta T_d = 10^{-5}$ K/√Hz are assumed, here just for purposes of comparison, for both cubic (configurations 1 and 2) and spherical GRS (configurations 3 and 4).

A comparison among acceleration disturbances to the TM that would be expected for all four configurations in a drag-free mode is presented hereafter.

## 4. Estimates of acceleration disturbances

One general classification of acceleration disturbances for a nearly free-falling object in a drag-free SC is shown in figure 6.

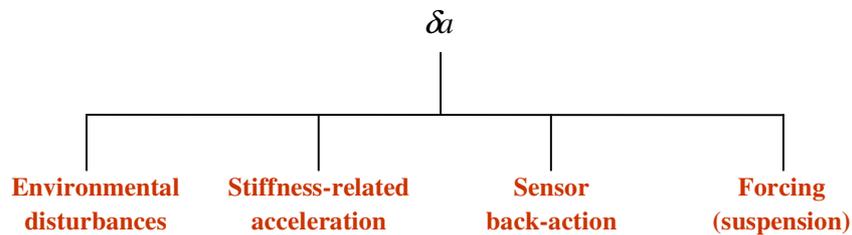

**Figure 6.** Classification of acceleration disturbances.

---

[9] other alloys such as a Cu-Be alloy would guarantee extremely low susceptibility to magnetic fields of $\chi \sim 10^{-6}$, and adequately high values of density [39].

**Table 1.** Main parameters of the four different configurations. Highlighted sections are where differences exist between the cubic and spherical GRS designs.

| | Symbol | Cubic GRS Configurations 1/2 | Spherical GRS Configurations 3/4 |
|---|---|---|---|
| Spacecraft mass | $m_{SC}$ | 300 kg | |
| Spacecraft surface area | $A_{eq}$ | 4 m$^2$ | |
| Orbital velocity | $v_{orbit}$ | 29.78 km/s | |
| Solar irradiance | $W_o$ | 1360 W m$^{-2}$ | |
| Variance of solar irradiance | $\delta W_o / W_o$ | $1.3 \cdot 10^{-3}$ $1/\sqrt{Hz} \cdot \left(\frac{1\,mHz}{f}\right)^{1/3}$ | |
| Mass of the TM | $m_{TM}$ | 1.9 kg | |
| Volume of the TM | $V_{TM}$ | $9.73 \cdot 10^{-5}$ m$^3$ (L = 46 mm, cube; r = 2.85 cm, sphere) | |
| Density of TM | $\rho_{TM}$ | $1.96 \cdot 10^4$ kg m$^{-3}$ | |
| Equivalent TM area | $\alpha_{(i)}$ | (dep. on physical effect (i), see perf. tables, section 4) | |
| Gap between TM and housing | $d$ | $d$ = 4 mm (along meas. axis) | $d$ = 20 mm (spherical housing) |
| TM magnetic susceptibility | $\chi$ | $1.7 \cdot 10^{-5}$ | |
| TM remnant mag. moment | $M_r$ | 20 nA·m$^2$ | |
| Local magnetic field: DC, fluctuation | $B_i^{DC}$, $B_i$ | $10^{-5}$ T , 50 nT/$\sqrt{Hz}$ | |
| Local mag. field gradient | $\left(\frac{\partial B_i}{\partial x_j}\right)^{DC}, \left(\frac{\partial B_i}{\partial x_j}\right)$ | $5 \cdot 10^{-6}$ T·m$^{-1}$, $25 \cdot 10^{-9}$ T m$^{-1}$/$\sqrt{Hz}$ | |
| DC mag. second derivative | | 0.2 T·m$^{-2}$ | |
| Interplanetary mag. field fluctuation | $\delta B_{int}$ | 40 nT/$\sqrt{Hz}$ at 1 mHz, rising as steeply as 1/f at low frequency (Ulysses data) | |
| Magnetic shielding factor | $\xi$ | $10^{-2}$ | |
| TM charge | $q$ | $1.6 \cdot 10^{-13}$ C | |
| TM charge fluctuation | $\delta q$ | $\sim (2e\dot{q})^{1/2}/2\pi f \sim 5.85 \cdot 10^{-16} \cdot \left(\frac{1\,mHz}{f}\right)$ C/$\sqrt{Hz}$, $\dot{q}$ = 260 $e \cdot s^{-1}$ | |
| Stray DC electrode potential | $V_p$ | 100 mV | |
| Voltage fluctuation Capacitive sensing | $\delta V$ | 10 µV/$\sqrt{Hz}$ see table 4 | --- full optical sensing |
| Pressure inside the housing | P | 10 µPa | |
| Mass of residual gas | $m_{gas}$ | $6.69 \cdot 10^{-26}$ kg | |
| TM/housing temperature | T | 293 K | |
| Temperature difference fluctuation | $\delta T_d$ | $3.7 \cdot 10^{-6}$ K/$\sqrt{Hz}$ (at $10^{-4}$ Hz) | |
| Incident laser power | $P_{laser}$ | 0.1 mW | |
| Relative laser power fluctuation | $\delta P_{laser}$ | $10^{-4}$ $1/\sqrt{Hz}$ | |

The total deviation of the TM from a geodetic motion, i.e. the total (non-gravitational) acceleration noise $\delta a$ acting on the TM, would result from: environmental disturbances $\delta a_{env}$, acceleration noise from stiffness $\delta a_{stiff}$, sensor-back actions $\delta a_{ba}$, and TM suspension (forcing) $\delta a_f$, whenever suspension is required. The contribution of each of these noise sources to $\delta a$ is estimated in the remainder of the paper for the four configurations in a LISA-like noise environment.

4.1. *Environmental disturbances*

Environmental disturbances is acceleration noise due to: *magnetic disturbances*, i.e. fluctuations of local magnetic field and interplanetary magnetic field; *self-gravity disturbances*, i.e. fluctuations of the temperature field on the surrounding SC which would cause thermo-elastic distortions of SC and payload and result in self-gravity actions on the TM; *thermal disturbances,* i.e. fluctuations of temperature differences across the GRS which would result in differential gas pressure (radiometer effect), differential radiation pressure, and asymmetric out-gassing; *laser power noise*, i.e. fluctuating laser radiation pressure on the TM; collision impacts from *cosmic rays*, and *residual gas* (Brownian noise).

The physical sources of these disturbances are common to all configurations. For configurations 1 and 2, these disturbances are estimated largely based on the model reported in table 3a of [13] using design parameters summarized in table 1 above. An adaptation of that model has been used to estimate those same disturbances that would act on a spinning / non-spinning sphere.

In addition to disturbances that are common to all configurations, some other sources of noise would be peculiar of a spinning spherical TM; these are addressed as "spinning-sphere-only" accelerations in table 2 and estimated in some detail below.

Calculations in table 2 refer to 1 mHz, values are in [$10^{-16}$ m s$^{-2}$/√Hz] units, noise values smaller than 1% ($10^{-17}$ m s$^{-2}$/√Hz) of the total acceleration noise budget are not reported in the table and just neglected.

Acceleration disturbances that are identified and taken into consideration as particular to a spinning sphere are:

1) Interaction of the additional magnetic moment due to a charged spinning sphere with the residual magnetic field inside the GRS: a residual charge on a spinning TM would be the source of an additional magnetic moment that would interact with the local magnetic field to produce acceleration disturbances acting on the TM. It is straightforward to prove that even for $10^{-12}$ C charge on the TM, and a spin frequency of 10 Hz, the magnetic moment that would be due to spinning is on the order of $M_{spin} \sim 10^{-4} M_r$, i.e. four orders of magnitudes smaller than the residual magnetization expected for the TM. Resultant acceleration noise is therefore well below 1% of the total budget, and is not reported in table 2.

2) Interaction of Barnett moment [41], [49] with residual magnetic field inside the housing: even in the absence of external fields, if a TM with a small residual magnetic susceptibility is spun up at a given angular velocity, a precession of local magnetic moments inside the TM (that would have the same sense no matter what the orientation of the local moments) would occur. That would result in a net effective current (in the same sense of the angular velocity) which would couple to the residual magnetic field inside the GRS to produce acceleration noise. Calculations have shown that this acceleration noise would be significantly smaller than the one coming from the interaction between the residual magnetization of the TM and the local magnetic field, and, in any case, below 1% of the total budget.

3) Interaction of "low-frequency" attitude motion of the TM in the housing (due to constellation dynamics) with "high-order" harmonics of patch potential: the TM would be spun up orthogonally to the ecliptic plane, and its spin axis would tend to remain "very stably"[10] oriented in the inertial space during science runs. Since the constellation rotates in its plane once per year and the constellation plane precesses around the ecliptic pole once per year, the attitude of the TM would be changing inside the GRS over one year period. For an observer sitting on one LISA SC, this would correspond to a drift of the spin axis of the TM in the housing frame on the order of 0.2 μrad/s [31]. This "very-low-frequency" attitude variation of the TM inside the GRS could couple to "high-order" patch fields inside the housing to produce acceleration disturbances within LISA band. It can be proven [40] that, for 0.2 μrad/s attitude variation, acceleration disturbances in the LISA band would result from 1 mm scale (and smaller) patch fields inside the GRS; but, for a gap of $d \sim 20$ mm between the TM and the housing walls, 1 mm-scale patches (that would decay exponentially with the gap [42]) would produce negligible acceleration on the TM.

## 4.2. *Stiffness-related acceleration*

In the presence of residual coupling (stiffness) between TM and surrounding SC, any TM-to-SC relative jitter is the source of direct acceleration disturbances $\delta a_{stiff}$, acting on the TM, which causes a deviation from a perfect geodetic motion. In other words, due to finite stiffness, the TM is disturbed from its geodesics as a result of: thrusters noise, GRS readout errors propagated through the drag-free control loop, disturbances on the SC that are insufficiently suppressed by control gain. For the baseline sensor, in the most general form, $\delta a_{stiff}$, i.e. projection of acceleration disturbances (due to stiffness) in the direction of one line of sight (LoS), can be written as:

$$\delta a_{stiff} = k\delta x + k_{xy}\delta y + k_{xz}\delta z + k_{x\theta}\delta\theta + k_{x\eta}\delta\eta + k_{x\varphi}\delta\varphi \qquad (4)$$

$k^m$ is magnetic stiffness, i.e. contribution to coupling due to local magnetic field gradients, and second derivatives of local magnetic field; $k^{sg}$ is stiffness from self-gravity gradient; $k^e$ is electric stiffness ($k^e = k^{ic} + k^v + k^{av} + k^{pf}$), i.e. due to image charges $k^{ic}$, local DC voltages in the sensor $k^v$, applied voltages $k^{av}$, patch fields $k^{pf}$ [13].

For configurations 1 and 2 (table 1), the total diagonal coupling $k$ is unlikely to be smaller than $k = k^m + k^{sg} + k^e = 4 \cdot 10^{-7}$ s$^{-2}$ [13]; neglecting non diagonal stiffness for the purpose of the present analysis / comparison, with a relative jitter of $\delta x = 0.32$ nm/√Hz at 1 mHz for configuration 1, that would result from the design in [43][11], and with a relative jitter of $\delta x = 0.29$ nm/√Hz at 1 mHz for configuration 2 [46], total acceleration disturbances (due to stiffness) of:

$$\delta a_{stiff} \approx k\delta x \approx 1.26 \cdot 10^{-16} \text{ m s}^{-2}/\sqrt{Hz}$$
$$\delta a_{stiff} \approx k\delta x \approx 1.17 \cdot 10^{-16} \text{ m s}^{-2}/\sqrt{Hz} \qquad (5)$$

at 1 mHz would perturb each cubical TM from its geodetic motion in configurations 1 and 2, respectively.

---

[10] B. Lange [22] has estimated that, a spherical TM spun up orthogonally to the ecliptic plane, in a LISA-like disturbance environment, would experience a drift of its spin axis (due to local zero-frequency torques) that might be as small as 0.6 μrad/yr.

[11] assuming that optical readout in the measurement direction is used for control.

Table 2a. Thermal and magnetic environmental disturbances for each of the four configurations.

| **Thermal and Magnetic Environmental Disturbances ($\delta a_{env}$)** | **2 cubes:** [$10^{-16}$ m s$^{-2}$/√Hz], 1 mHz | **1 cube:** [$10^{-16}$ m s$^{-2}$/√Hz], 1 mHz | **1 sphere:** (spinning) [$10^{-16}$ m s$^{-2}$/√Hz], 1 mHz | **1 sphere:** (no spin) [$10^{-16}$ m s$^{-2}$/√Hz], 1 mHz |
|---|---|---|---|---|
| **Magnetic Disturbances** | | | | |
| On-board magnetic field: Magnetic field fluctuation | \multicolumn{4}{c}{$\frac{2\chi V_{TM}}{m_{TM}\mu_0} B_i \left(\frac{\partial B_i}{\partial x_j}\right)^{DC}$ ~ 1.16} | | | |
| Magnetic field gradient fluctuation | \multicolumn{4}{c}{$\frac{1}{m_{TM}}\left(M_r + \frac{2\chi V_{TM}}{\mu_0} B_i^{DC}\right)\left(\frac{\partial B_i}{\partial x_j}\right)^{DC}$ ~ 5.4} | | | |
| AC noise (down converted) | \multicolumn{4}{c}{$\frac{2\chi V_{TM}}{m_{TM}\mu_0} B_i \left(\frac{\partial B_i}{\partial x_j}\right)$ ~ 3.65} | | | |
| Interplanetary magnetic field: Lorentz acceleration on the TM | \multicolumn{4}{c}{$\xi \frac{q}{m_{TM}} v_{orbit} \delta B_{int}$ ~ 0.05} | | | |
| Field fluctuation | \multicolumn{4}{c}{$\frac{2\chi V_{TM}}{m_{TM}\mu_0} \delta B_{int} \left(\frac{\partial B_i}{\partial x_j}\right)^{DC}$ ~ 2.33} | | | |
| **Total magnetic disturbances (rss)** | \multicolumn{4}{c}{**7.01**} | | | |
| **Thermal disturbances** | | | | |
| Radiometer Effect | $\frac{\alpha_t P}{2 m_{TM} T} \delta T_d$ ~ 0.8 | | $\frac{\alpha_t P}{2 m_{TM} T} \delta T_d$ ~ 0.42 | |
| Radiation pressure asymmetry | $\frac{\alpha_p \sigma}{2 m_{TM} c} T^3 \delta T_d$ ~ 4.1 | | $\frac{\alpha_p \sigma}{2 m_{TM} c} T^3 \delta T_d$ ~ 2.4 | |
| Asymmetric out-gassing | $\frac{\alpha_o l_o \theta_o}{2 m_{TM} T^2} \delta T_d$ ~ 1.5 | | $\frac{\alpha_o l_o \theta_o}{2 m_{TM} T^2} \delta T_d$ ~ 1.5 | |
| **Total thermal disturbances (rss)** | **6.4** | | **3.7** | |

**Table 2b.** Total environmental disturbances for each of the four configurations.

| Environmental Disturbances ($\delta a_{env}$) | 2 cubes: [$10^{-16}$ m s$^{-2}$/√Hz], 1 mHz | 1 cube: [$10^{-16}$ m s$^{-2}$/√Hz], 1 mHz | 1 sphere: (spinning) [$10^{-16}$ m s$^{-2}$/√Hz], 1 mHz | 1 sphere: (no spin) [$10^{-16}$ m s$^{-2}$/√Hz], 1 mHz |
|---|---|---|---|---|
| **Magnetic Disturbances (rss)**[a] | 7.01 | | 7.01 | |
| **Thermal Disturbances (rss)**[a] | 6.4 | | 3.7 | |
| **Self-gravity disturbances (rss)**[d] | 1.1 | | 1.1 | |
| **Laser power noise** | $\frac{2\delta P_{laser}}{m_{TM}c} \sim 2.4$ | | $\frac{2\delta P_{laser}}{m_{TM}c} \sim 2.4$ [c] | |
| **Cosmic Rays** | negligible[b] | | negligible[b] | |
| **Others (Brownian noise):** | | | | |
| Residual gas | $\left(\frac{P\alpha_t}{m_{TM}}\right)^{1/2}(3k_b Tm_{gas})^{1/4} \sim 3.1$ | | $\left(\frac{P\alpha_t}{m_{TM}}\right)^{1/2}(3k_b Tm_{gas})^{1/4} \sim 1.5$ | |
| Dielectric losses | $\left(\frac{\sqrt{2}V_0 C}{m_{TM}} + \frac{q}{6m_{TM}d}\right)\sqrt{\frac{4k_b T\delta_{sens}}{\pi fC}} \sim 1.16$ | | negligible[b] | |
| **Sphere only terms**[e] | --- | | negligible[b] | |
| **Total environmental disturbances (rss)** | 10.7 | 10.7 | 8.49 | 8.49 |

[a] see table 2a.
[b] negligible is defined as ($\delta a < 10^{-17}$ m s$^{-2}$/√Hz at 1 mHz).
[c] negligible for "balanced" ORO design.
[d] self-gravity disturbances result from SC thermo-elastic distortions, $2GM_{dis}x^{-2} \cdot CTE \cdot \delta T_{SC}$, and similarly from GRS distortions, $\delta T_{SC} \approx 2 \cdot 10^{-3}$ K /√Hz at 1 mHz, CTE aluminium = $2.5 \cdot 10^{-5}$ K$^{-1}$.
[e] sphere only effects taken into account are:
   1) interaction of the additional magnetic moment due to a charged spinning sphere with a magnetic field in the housing;
   2) interaction of the Barnett moment with a magnetic field inside the housing [41], [49];
   3) interaction of low frequency motion of the TM orientation with respect to the housing (due to constellation dynamics) with high-order harmonics of patch potentials.

For configurations 3 and 4, $\delta a_{stiff}$ can be written as:

$$\delta a_{stiff} = \delta \underline{a} \cdot \underline{n}(\alpha) \qquad (6)$$

where $\underline{n}(\alpha)$ is the unit vector that identifies a line of sight to a distant spacecraft in the constellation, $\alpha$ is the constellation angle. Neglecting any stiffness crosstalk:

$$\delta a_{stiff} = k \cdot \sqrt{(\delta x \cdot \cos(\alpha/2))^2 + (\delta y \cdot \sin(\alpha/2))^2} \qquad (7)$$

where $\delta x$ and $\delta y$ are relative displacements along the two directions shown below in figure 7 and $k$ is diagonal stiffness, $k = k^m + k^{sg} + k^e$. Since:

$$k^e = k^{ic} + k^v + k^{pf}$$
$$k^{ic} \sim q^2 d^{-1} a_p^{-1}; \quad k^v \sim qV_{og} d^{-2}; \quad k^{pf} \sim a_p V_{pe}^2 d^{-3} \qquad (8)$$

and,

$$k^{sg} \sim \frac{2GM_{dis}}{r^3} \qquad (9)$$

for the spherical GRS design (configuration 3 and 4), a large gap (20 mm) would tend to reduce both $k^e$, and $k^{sg}$, thus producing a total diagonal coupling that might realistically be as small as $k = k^m + k^{sg} + k^e = 5 \cdot 10^{-8}$ s$^{-2}$ (right-side column in table 3).

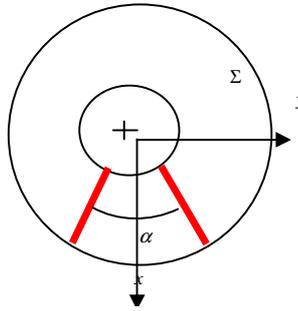

**Figure 7.** Housing reference coordinate system (spherical GRS design), *x-y* identifies the constellation plane.

For configuration 3, in the presence of the stiffness coupling above, and with:

- environmental disturbances acting on the SC, driven by fluctuations of solar radiation pressure as large as $\delta a_{SC} \approx \frac{2\delta W_o}{c} \cdot A_{eq} \cdot \frac{1}{m_{SC}} \approx 1.6 \cdot 10^{-10}$ m s$^{-2}$/$\sqrt{\text{Hz}} \cdot \left(\frac{1 \text{ mHz}}{f}\right)^{1/3}$ [13];
- direct (environmental) disturbances to the TM of $\delta a_{env} = 8.49 \cdot 10^{-16}$ m s$^{-2}$/$\sqrt{\text{Hz}}$ (at 1 mHz, table 2);
- noise in µN-thrusters (used to keep the SC centered around the TM) that would cause force fluctuations on the SC of $10^{-7}$ N/$\sqrt{\text{Hz}}$ (in the mHz frequency range);
- GRS readout precision of 5 pm/$\sqrt{\text{Hz}}$ (in the mHz frequency range), assumed to be achievable in real-time during science runs with a full optical readout [32], a relative jitter of $\delta x = \delta y = 0.3$ nm/$\sqrt{\text{Hz}}$ at 1 mHz in the constellation plane can be realistically achieved with a straightforward control design [23].

As a result, total acceleration disturbances due to stiffness at 1 mHz, for configuration 3, would be:

$$\delta a_{stiff} \approx k \cdot \sqrt{(\delta x \cdot \cos(\alpha/2))^2 + (\delta y \cdot \sin(\alpha/2))^2} \approx 0.15 \cdot 10^{-16} \text{ m s}^{-2}/\sqrt{\text{Hz}} \qquad (10)$$

smaller than the corresponding acceleration expected for configurations 1 and 2 as shown in table 3.

With a stiffness coupling of $k = k^m + k^{sg} + k^e = 5 \cdot 10^{-8}$ s$^{-2}$, same as in configuration 3, and a relative jitter of $\delta l = \sqrt{(\delta x \cdot \cos(\alpha/2))^2 + (\delta y \cdot \sin(\alpha/2))^2} \approx 12$ nm/$\sqrt{\text{Hz}}$ at 1 mHz [23], in the direction of one line of sight, configuration 4 would feature an acceleration noise from stiffness of:

$$\delta a_{stiff} \approx k \delta x \approx 6 \cdot 10^{-16} \text{ m s}^{-2}/\sqrt{\text{Hz}} \qquad (11)$$

significantly larger than in configuration 3, due the fact that TM sensing, which is used for drag-free control, would be degraded by asphericity errors.

Table 3. Acceleration noise due to stiffness coupling between spacecraft and proof-mass.

| Stiffness Related Acceleration ($\delta a_{stiff}$) | 2 cubes: | 1 cube: | 1 sphere: (spinning) | 1 sphere: (no spin) |
|---|---|---|---|---|
| **Total Stiffness in LoS** | $k = k^m + k^{sg} + k^e$ [13] | | $k = k^m + k^{sg} + k^e$ | |
| Magnetic Stiffness | $k^m$ [13] | | $k^m$ [13] | |
| Gravitational Stiffness | $k^{sg} \sim \dfrac{2GM_{dis}}{r^3} \sim 5 \cdot 10^{-9}$ s$^{-2}$ | | $k^{sg} \sim \dfrac{2GM_{dis}}{r^3} \sim 5 \cdot 10^{-9}$ s$^{-2}$ | |
| Electrical Stiffness | $k^e = k^{ic} + k^v + k^{pf}$ [13] | | $k^e = k^{ic} + k^v + k^{pf}$ ([13], model) | |
| image charges | $k^{ic} \sim \dfrac{q^2}{d a_p}$ | | $k^{ic} \sim \dfrac{q^2}{d a_p}$ | |
| DC voltages | $k^v \sim \dfrac{q V_{og}}{d^2}$ | | $k^v \sim \dfrac{q V_{og}}{d^2}$ | |
| patch fields | $k^{pf} \sim \dfrac{a_p V_{pe}^2}{d^3}$ | | $k^{pf} \sim \dfrac{a_p V_{pe}^2}{d^3}$ | |
| **Total Stiffness**[a] | $k \approx 4 \cdot 10^{-7}$ s$^{-2}$ | | $k \approx 5 \cdot 10^{-8}$ s$^{-2}$ | |
| **Relative PM to S/C jitter ($\delta$ in LoS) [nm/√Hz, 1 mHz]** | 1.44 (electrostatic) 0.32 (optical) [47] | 0.29 [46] | 0.3 [23] | 12 [c] [23] |
| **Total acceleration noise from stiffness**[b] **[10$^{-16}$ m s$^{-2}$/√Hz, 1 mHz]** | 5.75 (electrostatic) 1.26 (optical) | 1.17 | 0.15 | 6.0 |

[a] the model in this table predicts a stiffness of ~$1 \cdot 10^{-8}$ s$^{-2}$ for the spherical configuration. Due to uncertainties in the estimates of the stiffness especially magnetic and electrostatic effects [13], a conservative total stiffness of $k \sim 5 \cdot 10^{-8}$ s$^{-2}$ is assumed for the sphere based designs.
[b] although a potentially relevant source of disturbances, acceleration noise from stiffness cross-talk is not taken into consideration for the purposes of the present analysis / comparison.
[c] sensor operated with optical readout degraded by asphericity errors; motion of the CoM extracted from "raw data" during on-ground data analysis.

### 4.3. Acceleration noise due to forcing and sensor back-action

Configurations 3 and 4 do not require forcing during science runs, and have negligible sensor back-action (optical sensing), table 4 below. Although suspension errors, for configurations 1 and 2, are very design-dependent, and might affect acceleration performance negligibly (by proper design, and in-flight calibration), considering payload configurations that would require no forcing at all, whenever that could be achieved with a feasible and robust design, would be an attractive option for future space-based detectors beyond LISA.

Table 4. Acceleration noise from sensor back-action and applied control forces.

| Forcing and Sensor back-action ($\delta a_f$ & $\delta a_{ba}$) | 2 cubes: [$10^{-16}$ m s$^{-2}$/√Hz] Frequency (Hz) | | | 1 cube: [$10^{-16}$ m s$^{-2}$/√Hz] Frequency (Hz) | | | 1 sphere: (spinning) [$10^{-16}$ m s$^{-2}$/√Hz] Frequency (Hz) | | | 1 sphere: (no spin) [$10^{-16}$ m s$^{-2}$/√Hz] Frequency (Hz) | | |
|---|---|---|---|---|---|---|---|---|---|---|---|---|
| | $10^{-5}$ | $10^{-4}$ | $10^{-3}$ | $10^{-5}$ | $10^{-4}$ | $10^{-3}$ | $10^{-5}$ | $10^{-4}$ | $10^{-3}$ | $10^{-5}$ | $10^{-4}$ | $10^{-3}$ |
| **Forcing ($\delta a_f$)** | | | | | | | no forcing | | | | | |
| Suspension Errors[a] | -- | 8.2 | 10.8 | -- | 1.96 | 2.15 | --- | | | | | |
| Quantization Errors | negligible[b] | | | negligible[b] | | | --- | | | | | |
| **Total forcing acceleration noise ($\delta a_f$) [1 mHz]** | **10.8** | | | **2.15** | | | **0.0** | | | | | |
| **Sensor back-action[d]** | Frequency (Hz) | | | Frequency (Hz) | | | negligible[b] (optical sensing) | | | | | |
| | $10^{-5}$ | $10^{-4}$ | $10^{-3}$ | $10^{-5}$ | $10^{-4}$ | $10^{-3}$ | | | | | | |
| $\delta V_d \cdot V_{og} \sim \frac{C_x}{m_{TM}d}\frac{C_g}{C}V_{og}\delta V_d$ | 2.7 | 2.7 | 2.7 | 2.7 | 2.7 | 2.7 | --- | | | | | |
| $\delta V_d \cdot q \sim \frac{1}{m_{TM}d}\frac{C_g}{C}q\delta V_d$ | 0.44 | 0.44 | 0.44 | 0.44 | 0.44 | 0.44 | --- | | | | | |
| $\delta q \cdot V_d \sim \frac{1}{m_{TM}d}\frac{C_x}{C}V_d\delta q$ [c] | 228 | 22.8 | 2.28 | 228 | 22.8 | 2.28 | --- | | | | | |
| $q \cdot \delta q \sim \frac{1}{m_{TM}d}\frac{C_x}{C^2}q\delta q$ | negligible[b] | | | negligible[b] | | | --- | | | | | |
| **Total sensor back-action (rss)** | 228 | 23.0 | 3.88 | 228 | 23.0 | 3.88 | --- | | | | | |
| **Total sensor back-action (rss) ($\delta a_{ba}$) [1 mHz]** | **3.88** | | | **3.88** | | | **0.0** | | | **0.0** | | |

[a] contribution of suspension actuation (cross-talk) to acceleration noise in LoS from design and closed loop simulations [46, 47]; note that no translational suspension is required for configuration 2.
[b] negligible is defined as ($\delta a < 10^{-17}$ m s$^{-2}$/√Hz at 1 mHz).
[c] values are obtained under the assumption that no continuous charge control is performed (only discontinuous TM discharging).
[d] capacitive sensor parameters taken from [13]: $V_d = |V_{x1} - V_{x2}|$ = voltage difference across opposite faces along the x-axis (and other axes); $\delta V_d$ = fluctuations in voltage difference across opposite faces; $V_o = (V_{x1} + V_{x2})/2$ = average voltage across opposite faces $\approx 0.1$ V; $V_{og} = V_o - V_g$; $C_g [V_g]$ = capacitance [voltage] to ground; $C_g \approx C_x$. $C_x = \varepsilon_o \alpha_p / d \approx 6$ pF; $C \approx 6C_x$; $C' \approx (C_x/d^2)\Delta d \approx 1.5$ pFm$^{-1}$ ($\Delta d / 1 \mu$m); $\Delta d = d_{x1} - d_{x2}$ = asymmetry in gap across opposite faces of the TM.

For the baseline sensor, i.e. configurations 1 and 2, at 1 mHz, fluctuations of voltage difference across opposite faces along the x-axis coupled to average voltage across opposite faces, $\delta V_d \cdot V_{og} \sim \frac{C_x}{m_{TM}d}\frac{C_g}{C}V_{og}\delta V_d$, would drive acceleration disturbances from sensor back-action.

Nevertheless, at lower frequencies ($10^{-4}$ Hz, and, possibly $10^{-5}$ Hz), $\delta q \cdot V_d \sim \frac{1}{m_{TM}d}\frac{C_x}{C}V_d\delta q$, i.e. fluctuations of the charge on the TM coupled to voltage difference across opposite faces, which is required for capacitive sensing, would rise up as $1/f$, in the absence of continuous charge control, thus becoming one of the dominant acceleration disturbances at $10^{-4}$ Hz. With an acceleration noise of $2.3 \cdot 10^{-15}$ m s$^{-2}$/√Hz at $10^{-4}$ Hz (obtained for $\dot{q} = 260$ e/s), charge fluctuations would prevent instrument performance from being extended below $10^{-4}$ Hz, as it might be targeted for LISA follow-on missions, without continuous active charge management.

### 4.4. Total acceleration noise

The total acceleration noise, the rss of all disturbances acting on the TM, for each configuration is summarized in table 5. Each single contribution to the total deviation from a geodetic motion is reported as well. Within the accuracy of the model used for the present analysis, configurations 2, 3, and 4 tend to approach an "environmental-noise-limited" design; for configuration 2, the contribution to total acceleration noise from suspension, $2.15 \cdot 10^{-16}$ m s$^{-2}$/√Hz at 1 mHz (only attitude suspension required for one-cube configuration) is below the un-avoidable environmental noise floor, thus making disturbances from suspension less critical than for configuration 1.

Table 5. Comparison of the total acceleration noise for all configurations.

| Total Acceleration Noise ($\delta a$) | 2 cubes: [$10^{-16}$ m s$^{-2}$/√Hz], 1 mHz | 1 cube: [$10^{-16}$ m s$^{-2}$/√Hz], 1 mHz | 1 sphere: (spinning) [$10^{-16}$ m s$^{-2}$/√Hz], 1 mHz | 1 sphere: (no spin) [$10^{-16}$ m s$^{-2}$/√Hz], 1 mHz |
|---|---|---|---|---|
| Environmental Noise ($\delta a_{env}$) (table 2) | 10.7 | 10.7 | 8.49 | 8.49 |
| Stiffness ($\delta a_{stiff}$) (table 3) | 5.75 (electrostatic) 1.26 (optical) | 1.17 | 0.15 | 6.0 |
| Forcing ($\delta a_f$) (table 4) | 10.8 | 2.15 | --- | --- |
| Sensor back-action ($\delta a_{ba}$) (table 4) | 3.88 | 3.88 | negligible [a] | negligible [a] |
| **Total RSS of all disturbances ($\delta a$)** | **16.7 (electrostatic) 15.74 (optical)** | **11.64** | **8.5** | **10.4** |

[a] negligible is defined as ($\delta a < 10^{-17}$ m s$^{-2}$/√Hz at 1 mHz).

Configuration 3 achieves a minimal disturbance environment, by using a spinning sphere and a large gap to eliminate all suspension forces and minimize the impact of electrostatic patch effects. Configuration 4 takes a similar approach, but instead takes advantage of the reduced stiffness to significantly relax requirements on the drag-free position error. The increased gap size of these configurations significantly relaxes requirements on charge management, preventing performance from significantly degrading below 1 mHz due to fluctuations in the charge on the proof-mass.

### 5.  Concluding remarks

Future drag-free missions require extremely challenging disturbance reduction; for LISA the total deviation from a perfect geodetic motion has to be such that $\delta a < 3 \cdot 10^{-15}$ m s$^{-2}$/√Hz from $10^{-4}$ Hz up to 1 Hz, and even more demanding requirements may come from missions beyond LISA (such as BBO). As a result, the question that needs to be addressed is: how low can local non-gravitational disturbances be pushed with a realistic design for future missions?

In the attempt of answering this question, and identifying guidelines for future concepts that might be attractive for missions beyond LISA, estimates presented in this paper would indicate that acceleration disturbances as small as $\delta a \approx 8.5 \cdot 10^{-16}$ m s$^{-2}$/√Hz at 1 mHz for a one-spinning-sphere configuration, and as small as $\delta a \approx 1.16 \cdot 10^{-15}$ m s$^{-2}$/√Hz at 1 mHz, for a one-cube configuration, might be achievable in a LISA-like noise environment, both configurations approaching an "environmental-noise-limited" design.

Comparisons in tables 2, 3 and 4 suggest that future drag-free missions beyond LISA would benefit from "non-invasive" disturbance reduction concepts featuring no forcing applied to the TM, larger gap between the TM and its housing, and sensors less vulnerable to charge-related noise, whenever all that might be achieved with a feasible and adequately robust design. With regard to detector sensitivity in the frequency range, where acceleration disturbances are expected to be dominant, configuration 3 and 4 might be attractive options, since they would implement a comparatively "less-invasive" disturbance reduction concept. Nevertheless, how to design for missions beyond LISA should result from complete trade-off studies, which should involve not only experiment performance at "low-frequency", but also: measurement performance, payload design complexity, overall system considerations, complexity of drag-free control, complexity of experiment operations, required calibration efforts, constellation reliability, technology readiness, and program risk management.

**Acknowledgements**

This research has been carried out at Stanford University, Hansen Experimental Physics Laboratory, and EADS Astrium, Science Missions and Systems, Friedrichshafen. The authors would like to acknowledge funding from the German Aerospace Center (Deutsches Zentrum für Luft- und Raumfahrt) within the program "Investigations of system performance using alternative payload concepts for LISA" (50OQ0701).